\documentclass[aps,rmp,amsmath,amssymb,longbibliography]{revtex4-1}

\usepackage{bm}
\usepackage{graphicx}
\usepackage{epstopdf}
\usepackage{amsmath}
\usepackage{subfigure}
\usepackage{xcolor}
\usepackage{geometry}
\setcitestyle{numbers}
\begin{document}

\title{Quantum Chemical Calculation of Molecules in Magnetic Field}
\author{Mihir Date}
\author{R.W.A Havenith}
\affiliation{Zernike Institute for Advanced Materials, University of Groningen,
Groingen, 9743 LG, The Netherlands}

\date{April 10, 2020}

\begin{abstract}

\end{abstract}


\maketitle

\tableofcontents{}

\section{Introduction}
\label{intro}
The early success of first-principles/\textit{ab initio} calculations has inspired physicists and chemists to use these methods for the theoretical validation of experimentally observed phenomena. Since the rise in the popularity of NMR measurements for structure determination, a considerable effort has been put to develop first-principles methods to calculate chemical shifts and shielding constants. Theories, leading to their implementation in commercial software, were formulated early in the 1970s and 1980s, primarily by P.Lazzeretti and R.Zanasi (and collaborators)\cite{Lazzeretti1976, Lazzeretti1980, Lazzeretti1995, Fowler1996}. Around the same time, there was a growing interest to study the molecular spectra in the atmospheres of white dwarfs and neutron stars. Hydrogen, Helium, carbon, etc in the atmosphere of such massive objects are subjected to strong magnetic fields of the order of 10$^{4-7}$T\cite{Ostriker1968, Ruderman1971,Ruder1984}.  The existing theories around that time treated external magnetic field as a weak perturbation, as in case of NMR and therefore computing the energy spectrum was convenient with perturbation theory. But for field strengths as high as 10$^{4-7}$T, the magnetic interactions are no longer weak and cannot be treated by perturbation theory. Theoretical calculations at the Hartree-Fock level were presented primarily for He and C by Neuhauser et al and Demeur et al\cite{Neuhauser1986, Demeur1994}, while a Density Functional Theory (DFT) based approach was proposed by Vignale  and Rasolt\cite{Vignale1987}.

At this point, it is important to categorize magnetic field strengths as ``weak" and ``strong". For the purpose of this review, we consider magnetic fields attainable in a laboratory for NMR experiments as ``weak" and those observed due to the white dwarfs and neutron stars as ``strong". A comprehensive review of theoretical approaches towards dealing with atoms and molecules in weak magnetic fields, specifically for NMR applications, is presented in \cite{Helgaker1999}. For the sake of completeness, we mention a few important theoretical aspects relevant to their application in commericially available softwares. \\
Our main interest here is to present theoretical methods used to study energy spectra of molecules in ultrahigh/strong magnetic fields. This review is intended to aid non-expert theoreticians and experimentalists to choose the most suitable computational method to complement their experimental observations. Therefore, the extent of mathematical rigor is limited. \\
The review is organized into two main sections, where the first section presents an overview of the theory of magnetic response of atoms and molecules in weak magnetic fields, and the second section discusses the same problem in case of strong magnetic fields. We consider some specific examples in the second section to compare different theories. In most cases, the energy spectrum of carbon atom is used as an example for its emphatic importance in the atmospheric molecular band of white dwarfs and neutron stars\cite{Jura2014, Suleimanov2014, Ho2009}. For the convenience of units, atomic units are used for magnetic field throughout this review, where 1 a.u.= 2.34$\times$10$^4$T.

\section{ATOMS AND MOELCULES IN WEAK MAGNETIC FIELDS}
\label{sec1}
\subsection{The NMR Spin Hamiltonian and the Gauge Origin Problem}
For most practical applications, NMR experiments are the situations where molecules are subjected to low field strengths.  A detailed review on \textit{ab initio} calculation of shielding constants and chemical shifts is presented in \cite{Helgaker1999}. We only summarize the conceptual framework, which is crucial to the quantum chemical calculations performed by commercially available softwares.\\
We start by writing the NMR Spin Hamiltonian
\begin{equation}
    \hat{H}= -\sum_K \gamma_K \hbar \vec{B}^T (1-\hat{\sigma_K})\hat{I}_K + \frac{1}{2}\sum_{K\neq L} \gamma_K \gamma_L \hat{I}^T \hbar^2 (\hat{D}_{KL}+\hat{K}_{KL})\cdot\hat{I}_L .
    \label{nmr_hamiltonian}
\end{equation}
Here, $\gamma_K$ is the gyromagnetic ratio, $\hat{I}_K$ is the nuclear spin operator, $\hat{\sigma_K}$ is the magnetic shielding tensor (due to electron density), $\hat{D}_{KL}$ and $\hat{K}_{KL}$ are dipolar and indirect coupling constants. The notation of \cite{Helgaker1999} has been preserved for the reader's convenience. The isotropic Hamiltonian can be written as the rotational average of Eq.\ref{nmr_hamiltonian},
\begin{equation}
     \hat{H}= -\sum_K \gamma_K \hbar(1-\sigma_K)B\hat{I}_{K_z} + \frac{1}{2}\sum_{K\neq L} \gamma_K \gamma_L K_{KL}\hbar^2\hat{I}_K \cdot\hat{I}_L .
\end{equation}
where $\sigma_K= \frac{1}{3}tr(\hat{\sigma_K}$) and $K_{KL}=\frac{1}{3}tr(\hat{K}_{KL})$. The second term in the Hamiltonian can be compactly written in terms of the spin-spin coupling tensor \textbf{J}$_\textbf{{KL}}$, which is defined as
\begin{equation}
    \textbf{J}_{\textbf{KL}}= \frac{\hbar\gamma_K\gamma_L\textbf{K}_\textbf{KL}}{2\pi}.
\end{equation}
We then note that the shielding and indirect spin-spin coupling is four orders weaker than Zeeman and Dipolar Interactions and hence could be treated as first order perturbation. To find the NMR parameters from the Spin Hamiltonian, the energy (E($\vec{B},\vec{M}$)) is expanded in terms of magnetic field and magnetic dipole moment $\vec{M}$. This gives us the following relations, which are solved by variational perturbation theory.
\begin{equation}
    \boldsymbol{\sigma}_\textbf{K}= \vec{B}^T  \frac{d^2 E(\vec{B},\vec{M})}{d\vec{B} d\vec{M}_K} \vec{M}_K, \hspace{0.5cm} \hat{K}_{KL} + \hat{D}_{KL} = \hat{M}_L^T  \frac{d^2 E(\vec{B},\vec{M})}{d\hat{M}_L d\hat{M}_K}  \hat{M}_K
\end{equation}
It is a common practice to write the magnetic field as a curl of a vector potential, or a gauge potential. For a net magnetic field, including the contribution from the individual dipole moments and the external field, one can associate a total vector potential \textbf{A}$_\textbf{tot}$,
\begin{equation}
    \textbf{A}_{\textbf{tot}}= \textbf{A}_{\textbf{O}}(\vec{r}_i) + \sum_k \textbf{A}_k(\vec{r}_i)\,  \hspace{0.5cm} \vec{B}_{tot}= \nabla \times \textbf{A}_{\textbf{tot}}
\end{equation}
where the second term in summation is the contribution from individual magnetic dipole moment, and $\textbf{A}_{\textbf{O}}(\vec{r}_i)$ is the vector potential associated with the external magnetic field defined as 
\begin{equation}
    \textbf{A}_{\textbf{O}}(\vec{r})= \frac{1}{2} \vec{B} \times \vec{r}_{iO}
\end{equation}
The subscript of \textbf{A} denotes gauge origin and the notation $\vec{r}_{iO}=\vec{r}_i-\vec{r}_O$ is adopted. Although the NMR spectra depends largely on the spin Hamiltonian in Eq.\ref{nmr_hamiltonian}, electronic effects cannot be ignored. For an electron, it interacts with the vector potential due to the external magnetic field as well as the intrinsic field due to the nuclear magnetic moment. Thus, in general, the canonical momentum operator for an electron in magnetic field could be written as
\begin{equation}
    \boldsymbol{\pi}= \textbf{p}+q\textbf{A}
\end{equation}
Hence, in the Hamiltonian, terms linear in \textbf{A} and quadratic in \textbf{A} are obtained. By inserting the definition of \textbf{A}, we get
\begin{equation}
    \textbf{A}\cdot\textbf{p}= \frac{1}{2}(\vec{B} \times \vec{L}_O) , \hspace{0.4cm} \frac{1}{2}\textbf{A}^2= (\frac{1}{2} \vec{B}_{ext}\times\vec{r}_{iO})^2= \vec{B}\textbf{P}^{\zeta}\vec{B}
\end{equation}
here, $\textbf{P}^{\zeta}=\frac{1}{8}(\vec{r}^T_{iO}\vec{r}_{iO}\textbf{1}-\vec{r}_{iO}\vec{r}^T_{iO}$). The symbol $\zeta$ represents the type of nuclear-nuclear or nuclear-electronic interaction. The magnetic field due to the nuclear spin moment, which interacts with the electron spin is given as
\begin{equation}
    \vec{B}(\vec{r})= \frac{-g_K\mu_N}{c^2}[\frac{\hat{I}_K}{|\vec{r}-\vec{R}_K|^3}-\frac{3(\vec{r}-\vec{R}_K)((\vec{r}-\vec{R}_K)\cdot\hat{I}_K}{|\vec{r}-\vec{R}_K|^5}-\frac{8\pi}{3}\hat{I}_K\delta_{\vec{r},\vec{R}_K}]
\end{equation}
where, $g_K$ is the nuclear g-factor and $\mu_N$ is the nuclear magnetic moment This field interacts with the electron through $g_e\mu_B\boldsymbol{s}\cdot \vec{B}$ and gives rise to the following set of nuclear-electronic (ne) and nuclear-nuclear(nn) interactions:
\begin{equation}
    \hat{H}_{ne}^{SD}= g_eg_K\mu_B\mu_N \boldsymbol{s}\cdot \frac{(\vec{r}^T_{iO}\vec{r}_{iO}\textbf{1}-\vec{r}_{iO}\vec{r}^T_{iO})}{\vec{r}_{iO}^5} \hspace{0.5cm}       \text{Spin-Dipolar (SD)}
\end{equation}
\begin{equation}
    \hat{H}_{ne}^{FC}= \boldsymbol{s}\cdot\hat{I}_K \delta_{\vec{r},\vec{R}} \frac{8\pi g_Kg_e\mu_N\mu_B}{3c^2} \hspace{0.5cm} \text{Fermi-Contact(FC) operator}
\end{equation}
\begin{equation}
    \hat{H}_{ne}^{PSO}= (\frac{g_K\mu_N}{c^2})\frac{\hat{I}_K \cdot\vec{L}_{O}}{|\vec{r}_{iO}|^3} \hspace{0.5cm} \text{Paramaganetic Spin Operator (PSO)}.
\end{equation}
In the presence of an external magnetic field, the diamagnetic coupling (linear in $\vec{B}_{ext}$) of the external magnetic field, with an associated gauge centered at \textbf{G}, is given by 
\begin{equation}
    \hat{H}^{DS}_{nn}= \frac{g_K\mu_N}{2c^2}\vec{B}_{ext} \frac{\vec{r}_{iG}^T r_{iO}\textbf{1}-\vec{r}_{iO} r_{iG}^T}{|r_{iO}|^3} \hspace{0.5cm} \text{Nuclear Diamagnetic Shielding operator(DS)}
\end{equation}
One can notice that the interaction terms, which scale as $c^{-2}$ could be treated as weak perturbations and thus we obtain the shielding tensor by correcting Diamagnetic Shielding (DS) operator to the second order
\begin{equation}
    \boldsymbol{\sigma}_K= \langle\phi_{HF}|\hat{H}_{ne}^{DS}|\phi_{HF}\rangle -2\sum_{i\neq0} \frac{\langle\phi_{HF}|\hat{L}_{iO}|\phi_i\rangle-\langle\phi_{HF}|\hat{H}^{PSO}_{ne}|\phi_{HF}\rangle}{E_{HF}-E_i}
    \label{ramsey}
\end{equation}
The above expression is due to Ramsey\cite{Ramsey1951}. Here the magnetic field strength is a few Teslas, but when the molecules are in strong magnetic fields, the perturbation cannot be treated as ``weak". This problem is discussed in Section\ref{strong_field_sec}.

From Eq.\ref{ramsey}, the shielding tensor depends on $\hat{H}_{ne}^{DS}$ and $\vec{L}_{iO}$, which depends on the choice of gauge origin.
We notice that the shielding tensor (and chemical shift) should not depend on the coordinates of the molecule in space, that is, it should be gauge independent. This can be assured when the calculation is performed with a complete basis set. However, for practical reason (computational cost and power), we often use a optimized basis set (determined through convergence tests). The first term in Eq.\ref{ramsey} is just the expectation value of and can be determined with a good precision. But in the second term, the effect of magnetic field may generate a function outside our defined basis set and hence, the accuracy of the second term is compromised. This can be illustrated by considering a basis set consisting of Gaussian Type Oribtals (GTO), with certain Gaussian centers. Further, let the gauge origin span complete set of Gaussian centers. In a complete basis, irrespective of the gauge origin, a function generated upon a gauge dependent operation necessarily lies in the basis. But in an optimized basis, the gauge origin might lie outside the set of Gaussian centers and hence a function outside the basis set could be generated.
This introduces the gauge origin problem. Choice of gauge is just a tool for convenient mathematical construction, but the observables are gauge invariant. We can explore the prescription for the gauge origin to be the nuclear position. From elementary gauge transformations, the vector potential transforms as
\begin{equation}
    \textbf{A}_\textbf{O'}(\vec{r})= \textbf{A}_{O}(\vec{r}) + \nabla\cdot f(\vec{r}), \hspace{0.5cm} \nabla \times \nabla\cdot f=0
    \label{gauge_trans}
\end{equation}
where $f(\vec{r})$ is a scalar function and \textbf{O'} is the position vector of the transformed origin. The conditions in Eg.\ref{gauge_trans} ensure gauge invariance. The wavefunction, in case of such a transformation, picks up a phase $e^{-if\cdot r}$, where $f(\vec{r})$ has the following definition.
\begin{equation}
    f(\vec{r})= \frac{1}{2}\vec{B}\times(\textbf{O'-O})
\end{equation}
So applying this to a one electron Slater orbital $\chi_{lm}$, we obtain an overall wavefunction, which is known as the Gauge Included Atomic Orbital (GIAO).
\begin{equation}
    \omega(\textbf{A}_\textbf{O})= e^{-\iota \frac{1}{2}\vec{B}\times(\textbf{N}-\textbf{O})}\chi_{lm}
    \label{giao}
\end{equation}
We shall encounter this scheme of writing atomic orbitals in future contexts, however the notation might differ slightly from Eq.\ref{giao}.\\

The chemical shift tensor is proportional to the shielding constant and can be defined as 
\begin{equation}
    \boldsymbol{\sigma'}(\vec{r})= -\vec{B}_{in}\vec{B}^T
\end{equation}
where, $\vec{B}_{in}$ is the induced magnetic field due to the ring current density $\textbf{j}(\vec{r}$), defined as
\begin{equation}
    \vec{B}_{in}= \frac{1}{c}\int d^3r' \textbf{j}(\vec{r'}) \times \frac{\vec{r}-\vec{r'}}{|\vec{r}-\vec{r'}|^3}
\end{equation}
The current density is the crucial component of these integrals and is treated in two fashions-the GIPAW method (Gauge Included Projector Augmented Wave) or the CTOCD method (Continuous Transformation of Current Density). The GIPAW method, explained by Pickard and Mauri \cite{Pickard2001} is used in the popular software packages CASTEP\cite{Clark2005, Joyce2007} and Wien2K\cite{Laskowski2012, Laskowski2014}. The CTOCD-DZ (DZ:Diamagnetic contribution set to zero)\cite{Keith1993, Lazzeretti1994}, described by Ligabue et al\cite{Ligabue2003}, is implemented in the software package DALTON\cite{Aidas2014}. It is beyond the limitations of this review, to condense these theories and to discuss their implementation. Nonetheless, we leave the discussion by remarking that these methods are primarily different in the choice of bases they use; the GIPAW method is suitable for plane wave basis set and therefore advised for the solid state, while the CTOCD-DZ uses Gaussian type basis and is more suitable for molecules. The reader is strongly advised to refer to the cited articles above for a detailed explanation of these methods.
\section{ATOMS AND MOLECULES IN STRONG MAGNETIC FIELDS}\label{strong_field_sec}
\subsection{Hartree-Fock based methods}
In the previous section, we treated the effect of external magnetic field as a weak perturbation. However, in the cases where the magnetic field strengths appear in the magnitudes of 10$^{4-7}$T, the problem could no longer be solved using perturbation theory.\\
We start our discussion with the simplest atom; the hydrogen atom. Kappes and Schmelcher \cite{Kappes1994} presented basis optimization of London type wavefunctions. For this, we begin by choosing out basis of the kind,
\begin{equation}
    \Psi(\textbf{r}, \alpha, \textbf{R}, \textbf{C})= e^{-\iota\textbf{A}\cdot\textbf{r}} (x-R_x)^{n_x}(y-R_y)^{n_y}(z-R_z)^{n_z} e^{-\alpha(\textbf{r}-\textbf{R})^2}
    \label{london_orbital}
\end{equation}
where, \textbf{C} is the position vector chosen as the gauge origin (otherwise spanned by position vector \textbf{R}), and \textbf{A(C)} is the vector potential at position \textbf{C}. The complex phase multiplied at the beginning promises gauge invariance. In our calcuations, $\alpha$ enters as a 3$\times$3 symmetric matrix (since \textbf{r-R} is a three component vector), which we shall use as a variational parameter for basis set optimization. \\
The external magnetic field could be chosen to be acting in the z-direction, which naturally conserves the z-component of the angular momentum. It is therefore, convenient to work with cylindrical coordinates. This modifies our basis set in the following fashion,
\begin{equation}
    \Phi_{k,l}^{m_\pi}= \rho^{m+2k}\cdot z^l \cdot e^{(-\alpha \rho^2-\beta z^2)} e^{\iota m \phi} 
    \label{basis_kappes}
\end{equation}
where, $\alpha$ and $\beta$ are the matrix elements, $\alpha_{xx}$=$\alpha_{yy}$= $\alpha$ and $\alpha_{zz}$=$\beta$. The Pauli Hamiltonian  matrix (see below) could then be constructed using the basis functions in Eq.\ref{basis_kappes} and then diagonalized to obtain eigenenergies.These eigenenergies are in terms of a given set of variational parameters {$\alpha_i, \beta_i$}. The ground state energy could be minimized with respect to these parameters iteratively. It is not of our interest to discuss optimization algorithm. A more useful result, which the authors reported, is the dependence of {$\alpha_i, \beta_i$} on the magnetic field. 
\begin{equation}
    \hat{H}_{Pauli}= \hat{H}_{el}-\frac{e}{m} \hat{L}\cdot\vec{B}+\frac{e^2}{8m}|\vec{B}\times \textbf{r}|^2
\end{equation}
Here, $H_{el}$ is the Hamiltonian of the unperturbed molecule. \\
It was observed by the authors that for low field strengths, $\alpha_i=\beta_i$, but they start to deviate from each other in the high field regime. This attributed to the different forms of orbital representation in the directions parallel and perpendicular to the field. In the parallel direction, the orbital dynamics is captured by Slater type representation $e^{-c |{z}|}$, while the in-plane orbital dynamics (perpendicular to the direction of field) could be described by the Landau orbitals\footnote{Landau Orbitals are functions that describe Landau levels in a symmetric gauge. It can be represented as $\Psi(x,y)=f(z)e^{c|z|^2}$, where $z=x+\iota y$}. 
This is a useful result to choose the type of orbitals we shall use as basis functions for our calculations. The authors report results on dissociation energies and bond lengths as a function of magnetic field strength (upto 1 a.u.), but we postpone the discussion for later sections. 
The London orbitals were further popularised by Tellgren et al \cite{Tellgren2009}
where they described non-perturbative formalism for quantum chemical calculations in magnetic field.\\
In their article, they describe the London orbitals in the same way as in Eq.\ref{london_orbital}, but with a slightly different interpretation. The phase factor we saw in Eq.\ref{london_orbital} appeared to addressed the well known gauge origin problem. But in this formalism, the phase is treated as a plane wave with the wavevector having the same form as that of \textbf{A}. 
Having said so much, we can attempt to formulate a two-electron problem, following the McMurchie-Davidson scheme\cite{McMurchie1978}.
Let us consider two single electron functions, described by Eq.\ref{london_orbital}, $\phi(\textbf{r,C})$ and $\phi(\textbf{r,D)}$, where \textbf{C} and \textbf{D} are position vectors of the gauge origin in each case. We can define a new ``center" or gauge origin for the overlap of these two orbital functions as
\begin{equation}
    \textbf{P=} \frac{c\textbf{C}+d\textbf{D}}{c+d}
\end{equation}
where, c and d are Gaussian coefficients of $\phi(\textbf{r,C})$ and $\phi(\textbf{r,D)}$ respectively, and $p=c+d$. The composite system of two function is described by the simple product rule 
\begin{equation}
    \phi(\textbf{r,C})\cdot\phi(\textbf{r,D}) = \Omega^{k_1}(\vec{r}_P)= e^{-\iota k_1\cdot r}\Omega(\vec{r}_P)
    \label{OD}
\end{equation}
where $k_1=k_C-k_D$.
Next, a magnetic field dependent spherical London orbital (known as the London Hermite-Gaussian wavefunction) is introduced, which describes the composite system. We obtain our energy expectation values in terms of this function.
\begin{equation}
    \Lambda_{tuv}^{k_1} (\vec{r}_P) = (\frac{\partial}{\partial P_x})^t  (\frac{\partial}{\partial P_y})^u  (\frac{\partial}{\partial P_z})^v e^{-p}\vec{r}_{P}^2 e^{-\iota k_1\cdot \vec{r}_P}
    \label{LHG}
\end{equation}
We notice that in Eq.\ref{LHG}, the function is secured from gauge origin problem as $k_1=\frac{1}{2} \vec{B}\times(C-D$) depends only on relative separation of Gaussian centers \textbf{C} and \textbf{D}.
We can write Eq.\ref{LHG} more compactly in the following form
\begin{equation}
     \Lambda_{tuv}^{k_1} (\vec{r}_P) = \Lambda_{tuv} (\vec{r}_P) e^{-\iota k_1\cdot \vec{r}_P}
     \label{fourier}
\end{equation}
Since the representations in Eq.\ref{OD} and Eq.\ref{LHG} describe the same system, their equivalence could be established by expanding $\Omega^{k_1}$ as shown below
\begin{equation}
    \Omega^{k_1}= \sum_{tuv} E_t^{n_x^c n_x^d} E_u^{n_y^c n_y^d} E_v^{n_z^c n_z^d} \Lambda_{tuv}^{k_1} (\vec{r}_P)
\end{equation}
where the expansion coefficients are field independent. It is then straightforward to show, from Eq.\ref{fourier} that 
\begin{equation}
    S^k_1= \sum_{tuv} E_{tuv}^{n_i^c n_i^d} \Lambda_{tuv}(k_1)
\end{equation}
where $\Lambda_{tuv}(k_1)$ is the Fourier transform of $ \Lambda_{tuv}^{k_1} (\vec{r}_P)$.

To describe the details of this formalism is beyond the scope of this review and hence we shall summarize the use of this aforementioned formalism by writing the Coulomb repulsion term in the following manner, for the composite system.
\begin{equation}
    (\phi_C\phi_D|\phi_F\phi_G) = \sum_{tuv}\sum_{\tau\nu\mu} E_{tuv}^{n_i^c n_i^d} E_{\tau\nu\mu}^{n_i^f n_i^g} ( \Lambda_{tuv}^{k_1} (\vec{r}_P)| \Lambda_{\tau\nu\mu}^{k_1} (\vec{r}_P))
    \label{coloumb_rep}
\end{equation}
The overlap of the London Hermite-Gaussian functions, in the summation, is computed as
\begin{equation}
    ( \Lambda_{tuv}^{k_1} (\vec{r}_P)| \Lambda_{\tau\nu\mu}^{k_1} (\vec{r}_P) = \frac{2\pi^{5/2}}{pq\sqrt{p+q}}  exp[\frac{k_1^2}{4p}] exp[\frac{k_2^2}{4q}]\times R_{tuv,\tau\nu\phi}^0
\end{equation}
where, $R_{tuv,\tau\nu\phi}^0$ is the zeroth order (n=0) function described by 
\begin{equation}
    R_{tuv,\tau\nu\phi}^n = \frac{\partial^{t+u+v}}{\partial P_x^t\partial P_y^u\partial P_z^v}\frac{\partial^{\tau+\nu+\phi}}{\partial Q_x^{\tau}\partial Q_y^{\nu}\partial Q_z^{\phi}}\times e^{-\iota\vec{k}_1\cdot\textbf{P}} e^{-\iota k_2\cdot\textbf{Q}}(-2\alpha)^n F_n(\alpha(\textbf{P'-Q'}))
\end{equation}
where, \textbf{P'}=\textbf{P}- $\frac{i}{2p}\vec{k}_1$, \textbf{Q'}=\textbf{Q}- $\frac{i}{2q}\vec{k}_2$, $\alpha= \frac{pq}{p+q}$ and $F_n$ is the nth-order Boys Function. The authors calculate the required $R^{0}_{tuv,\tau\nu\phi}$ using downward recursion (set of equations described in Eq.16 of Tellgren et al\cite{Tellgren2008}), where we start with $n^{th}$ order function (n$>$ 0) in the recursion and recover the zeroth order function. It is important to note that the effect of magnetic field only enters our calculation through the wavevectors $\vec{k}_1$ and $\vec{k}_2$. Hence, the interaction terms, like in Eq.\ref{coloumb_rep}, are a product of recursively solved molecular integrals, that include the effect of magnetic field. That is to say, this method does not treat the effect of magnetic field as a perturbation and therefore is a non-perturbative method. One can appreciate the fact that the theory presented above is suitable for any magnitude of the magnetic field. We rejoice over the fact that such non-perturbative treatment is appropriate to calculate molecular properties in strong magnetic fields.\\
This theory has been implemented at the Hartree-Fock level of computation at in the software package LONDON. The authors report their results on benzene, the HF molecule and a few diamagnetic systems, but we shall postpone our discussion on real molecules in the next section.\\
In our discussion  so far, we have not considered the structural consequences of subjecting molecules to strong magnetic fields. This problem is addressed by the authors of Tellgren et al\cite{Tellgren2012}, where they lay down the theory for computing molecular gradient that enables us to obtain the geometrically optimized structure in high magnetic fields. 

\begin{figure} [ht]
    \centering
    \includegraphics[width=0.5\textwidth]{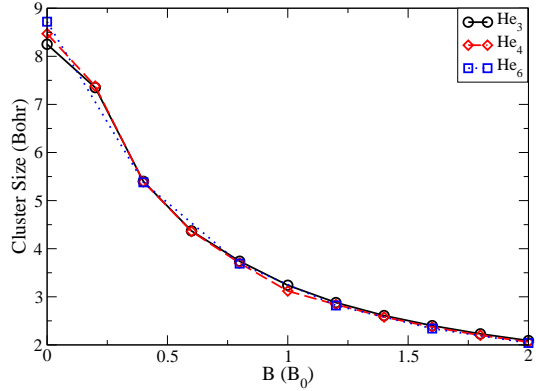}
    \caption{Contraction of cluster size with increasing magnetic field. The cluster size is measured as the smallest bond length in the structure with the optimized geometry. \textit{Reproduced from Tellgren et al, Phys. Chem. Chem. Phys., 2012, 14, 9492–9499}}
    \label{helium_cluster}
\end{figure}

We again start with the London type orbitals (they term it as Gaussian Type Orbitals-Plane Wave Orbitals, or GTO-PW orbitals), but now the wavefunctions are expressed in terms of a contracted basis
\begin{equation}
    \phi_\alpha(\vec{r})= \sum_{\textit{P}} \chi_p (\vec{r})U_\alpha^p
    \label{contracted_basis}
\end{equation}
where \textit{P} is the set of all primitive atomic orbitals and $\phi_\alpha$ is the set of real basis functions transformed, from the primitive basis, by the matrix $U_\alpha^p$. Another linear operator $\hat{L}$ (not to be confused with angular momentum operator) is introduced, which generates a set of primitive functions, which may not be contained in \textit{P}. This operation is defined as follows.
\begin{equation}
    \hat{L}\chi_p (\vec{r})= \sum_{\tilde{L}} v_q(\vec{r})L_p^q.
\end{equation}
The linear operator $\hat{L}$ acts on the real basis functions to yield the double summation 
\begin{equation}
    \hat{L}\phi_\alpha(\vec{r})= \sum_\textit{P} \sum_{\tilde{L}} v_q(\vec{r)}L_p^q U_\alpha^p
    \label{basis_trans}
\end{equation}
The matrix elements for this linear operation on basis functions, such that $\hat{L}\phi=\psi$, is given by
\begin{equation}
    L_{\alpha\beta} = \langle\phi_\alpha|\hat{L}|\phi_\beta\rangle= \langle\phi_\alpha|\psi_\beta\rangle 
\end{equation}
Now the authors consider $\hat{L_1}$ and $\hat{L_2}$ to be two linear operators mapping the space of primitive functions to space containing linear combination of those. One can represent an overlap integral/matrix element for an operator $\hat{\Omega}$ in the terms of non-primitive basis functions $\psi$ as
\begin{equation}
    \Omega_{\alpha\beta}= \langle\psi_\alpha|\hat{\Omega}|\psi_\beta\rangle
\end{equation}
The benefit of using non-primitive basis circumvents the memory issues in computing matrices of larger dimensions that we otherwise get while dealing with primitive basis. According to the authors, developing such a formalism aids encoding the quantum mechanics in object oriented programming languages (here, C++). \\
The authors treat the linear operator, mentioned above, as the partial derivative with respect to the nuclear coordinate \textbf{C}. They solve molecular gradients for the one dimensional case and employ it within RHF (Restricted Hartree Fock) theory to study Helium clusters. The molecular gradient for the one dimensional case is
\begin{equation}
    \frac{\partial E}{\partial C_x} = \frac{\partial D^{ji}}{\partial C_x} F_{ij} + \sum_i (D^{ji} F_{\frac{\partial_i}{\partial C_x},j} + c.c.)-\langle\chi_i|\sum_N \frac{\partial}{\partial C_x} \frac{Z_N}{|\vec{r}-\vec{R}_N|}|\chi_j\rangle D^{ji}
\end{equation}
where $D$ is the density matrix, whereas the elements $F_{ji}$ constitute the Fock matrix.
The authors studied He$_3$, He$_4$ and He$_6$ clusters, observed in the atmosphere of white dwarfs. To directly relate the shape of the clusters with the magnetic field strength is complicated, nonetheless, the cluster dimensions (measured as bond distances) reduce with increased magnetic field (see Fig.\ref{helium_cluster}). 

What causes such a contraction? Well, direct answers to the case of Helium clusters may not be available, but this is the appropriate juncture to discuss a distinguished bonding mechanism, known as the paramaganetic bonding, which we shall encounter in most of our discussions throughout this review.

Lange et al \cite{Lange2012} 
first proposed the paramagnetic bonding for the H$_2$ molecule, which they attribute to stabilization of anti-bonding orbitals perpendicular to the external magnetic field. In their calculations, authors perform FCI (Full Configuration Interaction) calculation, using an aug-cc-pVTZ uncontracted basis set, where the wavefunction is written as the anti-symmetric product of the spinors $\phi_p$ and the coefficients $C_n$ are obtained from Rayleigh-Ritz variational theory.
\begin{equation}
    |\Psi\rangle= \sum_n^N C_n det|\phi_{p1}\phi_{p2}\phi_{p3}...\phi_{pN}|
    \label{fci}
\end{equation}
This calculation is performed for GIAOs and therefore the set of equations discussed earlier are useful here as well.  \\
In case of a diatomic species, the authors distinguish two kinds of bonds- parallel to the applied field and perpendicular to the applied field. For the singlet case of the H$_2$ molecule, the parallel bonding is favored due to a greater overlap. The authors confirm this by plotting polar plots of potential energy surfaces in magnetic field (see Fig.1 of Ref.\cite{Lange2012}) .
\begin{figure}
    \centering
    \includegraphics[width=0.7\textwidth]{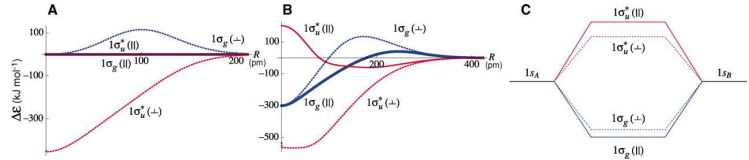}
    \caption{field induced change in bonding and anti-bonding orbital energies as a function of internuclear spacing, in parallel and perpendicular orientations of $H_2$ molecule. \textbf{a} is plotted by setting the exponential Gaussian coefficient a=1, whereas \textbf{b} is plotted with an optimized a. It is clearly evident that in both (a) and (b), the 1$\sigma_{u\|}^*$ energy curve lies above the 1$\sigma_{u\bot}^*$ curve. This is pictorially represented in the molecular orbital diagram in \textbf{c}. \textit{Reproduced with permission from \cite{Lange2012}.}}
    \label{lange_fig}
\end{figure}
The triplet, on the other hand, behaves conversely; electronic energy is lowered with the increased field strength, with preferred perpendicular orientation. To understand this stabilization mechanism, when the 1$\sigma_g$ molecular orbital is written as two 1s Gaussian orbitals, we realise that when the internuclear distance tends to zero, the MOs of the H$_2$ molecules could virtually be treated as atomic orbitals of He atom. 
\begin{equation}
1\sigma_{g/u} = [2\pm 2e^{-a/2 (1+B_x^2+B_y^2)R^2}]^{-1/2} \times (1s_A\pm1s_B)    
\end{equation}
Therefore, the authors prescribe that the 1$\sigma_g$ could be viewed as the 1s orbital of He, while the anti-bonding orbitals could be treated as $2p_0$ and $2p_{-1}$, with the former preferring parallel orientation and the latter prefers perpendicular orientation. It is found that the $2p_{-1}$ is stabilized than the $2p_0$, in strong magnetic field due to orbital Zeeman coupling. 
This hypothesis is verified by computing the bond energies for both the parallel and perpendicular orientations. Their plots clearly show that the 1$\sigma_{u\bot}^*$ is stabilized as compared to 1$\sigma_{u\|}^*$ (Fig.\ref{lange_fig}).

\subsection{Hartree-Fock methods for Helium and Beyond Helium}\label{examples_hf}
In this subsection, our aim is to discuss the examples where Hartree-Fock methods were employed, which significantly added to the conceptual framework of response of molecules in strong magnetic fields.\\
We go back in 1999 to discuss the results of Ivanov et al \cite{Ivanov1999}
, where they study the ground state of the carbon atom in strong magnetic fields. They perform calculations using the 2D Unrestricted Hartree-Fock method. They start by writing a simple expression for single particle energy of carbon atom in magnetic field as
\begin{equation}
    \epsilon_{B\mu}= (m_\mu+|m_\mu|+2S_{z\mu}+1)\cdot B/2-\epsilon_\mu
\end{equation}
where, $m_\mu$ is the magnetic quantum number, $S_{z\mu}$ is the z-projection of the spin $S_\mu$ and $\epsilon_\mu$ ia the one-electron energy. 

From their calculations, it is easy to identify configurations in the extremes-the field free and $B\rightarrow\infty$ case. The ground state (GS) configuration for the field-free case is $1s^22s^22p_0^12p_{-1}^1$ and it retains it form till a maximum field strength of 0.1862 a.u. For the latter case, the authors assume the fully polarized configurations, where each magnetic state (identified by the magnetic quantum number) contains one electron. Such a fully spin polarized configuration is given by $1s2p_{-1}3d_{-2}4f_{-3}5g_{-4}6h_{-5}$. While it is easy to identify these two ground states, it is difficult to identify the GS for intermediate field strengths. Hence, it was found that there exist multiple (5) ground states for the intermediate field strengths, which occur for the spin states $S_z= -1,-2,-3$. These observations are tabulated in Table\ref{mob:table1}.
\begin{table}[ht]
\caption{\label{mob:table1}The different atomic configurations of carbon under strong magnetic field (\textit{adapted from Ivanov et al\cite{Ivanov1999})}. We see that as the field strength increases, the system tries to maximise the magnitude of magnetic quantum number.}
\begin{tabular}{cc}
\hline \hline
\textbf{Magnetic Field Strength(a.u.)} &   \textbf{Configuration}\\
0-0.1862                               &   $1s^22s^22p_0^12p_{-1}^1$\\
0.1862-0.4903                          &   $1s^22s2p_0^12p_{-1}^12p_{+1}$\\
0.4903-4.207                           &   $1s^22s2p_0^12p_{-1}^13d_{-2}$\\
4.207-7.920                            &    $1s^22p_0^12p_{-1}^13d_{-2}4f_{-3}$\\
7.920-12.216                          &     $1s^22p_{-1}3d_{-2}4f_{-3}5g_{-4}$\\
12.216-18.664                         &     $1s2p_02p_{-1}3d_{-2}4f_{-3}5g_{-4}$\\
18.664-$\infty$                       &     $1s2p_{-1}3d_{-2}4f_{-3}5g_{-4}6h_{-5}$\\
\hline
\end{tabular}
\end{table}
The same formalism is extended, in their article in 2000\cite{Ivanov2000}
, to all the atoms from hydrogen to neon. Their results lay an important foundation for the results we shall discuss next in the context of these ions. In this article, they investigate the magnetic fields at which the transition from the fully spin polarized configuration happens to the field-free ground state. An important conclusion drawn from this study is that, for Z$\geq$6, there could be two competing fully spin polarized configurations, namely $1s2s2p_0$... and $1s2p_03d_{-1}...$. For lighter atoms, there exists a global fully spin polarized configuration, represented by$1s2p_{-1}3d_{-2}4f_{-3}5g_{4}6h_{-5}$.

\begin{figure} [ht]
    \centering
    \includegraphics[width=0.5\textwidth]{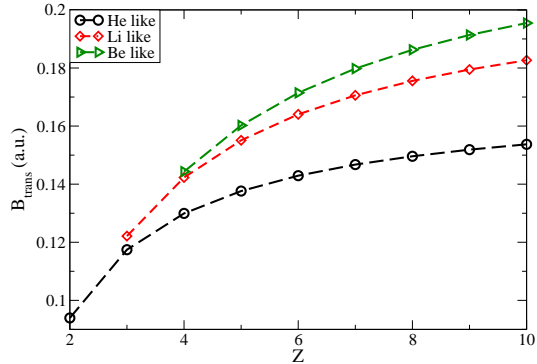}
    \caption{Variation in transition magnetic field from the all spins up ground state $|1s^\alpha...\rangle$ to the fully spin polarized state $|0\rangle$. \textit{Data reproduced from \cite{Boblest2014}.} }
    \label{boblest_magnetic_field}
\end{figure}

We now present a more modern Hartree-Fock based method, presented by Boblest et al \cite{Boblest2014} 
, which couples with Diffuse Quantum Monte Carlo (DQMC) to compute the energy spectrum of molecules in intense magnetic fields. In this method, authors write single particle spinors $\Psi^i$ in terms of Landau orbitals $\Phi_{i}$
\begin{equation}
    \Psi^i(\rho_i, \phi_i, z_i)= \sum_{n=0}^N P_n^i(z_i) \Phi_{nmi} (\rho_i,\phi_i)
\end{equation}
where, $P_n^i$ is expanded in terms of B-splines as follows
\begin{equation}
    P_n^i(z_i)= \sum_\nu \alpha_{n\nu}^i B_\nu^i (z_i)
\end{equation}
Here the B-spline coefficients $\alpha_\nu$ enter as variational parameter and minimization of the energy functional generates a set of equations known as the Hartee-Fock-Roothaan equations, which are solved iteratively. The energies obtained here are then improved by the DQMC, for which the importance sampled imaginary time Schr\"odinger Equation is solved. Here, the trial wavefunction is constructed simply by the forming a product of all spins up ($\psi_\alpha$) and all spins down ($\psi_\beta$) states, modulated by the Pade-Jastrow factor J\cite{Hammond1994}.
\begin{equation}
    \Psi_T(\vec{R})= J \psi_\alpha\cdot\psi_\beta
\end{equation}
The Pade-Jastrow factor is essentially optimized in the Monte-Carlo calculations. The energy is estimated by computing the local energy as $H\Psi_T/\Psi_T= E_L$.\\
The authors find that the previously reported configurations for the ground and fully polarized state, for 2 to 4 electron systems is consistent with their results. They also report variation of the transition magnetic field (magnetic field at which configuration transition occurs) with the atomic number Z. In such cases, as the Z number increases, the nuclear interactions overwhelm electron-electron interactions. This leads to a saturation of the transition magnetic field at about 0.17058 a.u.
Here, we introduce the following notation to denote the atomic configuration of atoms in different field strength regimes. The configuration of Fully Spin Polarized state $1s2p_{-1}3d_{-2}4f_{-3}5g_{-4}6h_{-5}$  is denoted by $|0\rangle$. Then the state $1s^22p_{-1}3d_{-2}4f_{-3}5g_{-4}$ is compactly written as $|1s^2...\rangle$, that is the ket contains the orbital that is different from the $|0\rangle$ configuration.
For He-like, Li-like and Be-like ions, the variation in the magnetic field strengths for the $|1s^\alpha...\rangle\rightarrow|0\rangle$ transition, with the atomic number is presented in Fig.\ref{boblest_magnetic_field}. 

The authors adopt a similar procedure for the Boron and Carbon like ions. For $C^+$, they determine that $|1s2p_0...\rangle$ is the ground state for field strengths lower than 0.0788 a.u. They also predict a possible transition path, for different magnetic field regimes, for 6-electron systems as $|1s2p_0...\rangle\rightarrow|1s...\rangle\rightarrow|2p_0...\rangle\rightarrow|0\rangle$. Their results for the neutral atom differ from those reported by Ivanov et al; they also find a $|2p_03d_{-1}...\rangle$ as a possible ground state. These discrepancy needs to be resolved by studying the fully spin polarised states in Neon by other methods including correlation or exchange interactions.

Following the aforementioned formalism of 2D HF,  Thirumalai et al \cite{Thirumalai2014} 
in 2014, solved the problem of the carbon in a slightly different fashion. The solution to single electronic hydrogenic system is obtained, without considering any exchange or direct interactions. The wavefunctions obtained are useful for two reasons; first, they work as initial estimates for the (coupled) Hartree-Fock problem and second, they are used to determine solutions to the elliptic partial differential equations (as described in \cite{Thirumalai2014}
, which give us direct and exchange potentials. The coupled Hartree-Fock problem is solved iteratively to extract eigenstates and eigenenergies.\\
They observe that there are twelve possible fully spin polarized configurations for the carbon atom, of which only two are reported by Ivanov et al, in their article in 2000. The authors claim that the key success of this method is its low computation cost at non-compromised accuracy. Hence, this method is also viable than FCI, which might end up being too expensive for a 6-electron system. However, the authors point out certain limitations of this method. Here, the relativistic effects are neglected, which might prove important for such high magnetic fields. Secondly, their equations treat the mass of nucleus as infinite in comparison with that of the electron. But in the case of ultrahigh magnetic fields, nuclear effects would also become relevant and influence of nuclear effects should be investigated. Lastly, as most Hartree-Fock methods do, this one also ignored electron-electron correlation effects.

\subsection{Rationalization of HF Results for Paramagnetic Closed Shell Systems}

It is well known that closed shell systems are expected to exhibit diamagnetic (induced magnetic field opposes external one) ground state in the zero field caase. However, works of Hegstrom and Lipscomb\cite{Hegstrom1966, Hegstrom1968, Hegstrom1968a} suggested the existence of molecules like BH (and isoelectronic compounds, due to Fowler and Steiner\cite{Fowler1991}) exhibit paramagnetism in the ground state, in the field-free case. On the contrary to the behaviour of conventional closed shell molecules, paramagnetic closed shell systems undergo a transition to a diamagnetic excited state at high magnetic fields.\\
Tellgren et al\cite{Tellgren2009} performed Hartree-Fock based calculations for such molecules in strong magnetic fields. An important output of their work is a simple two level model, which is used to rationalize and fit the calculated energy spectrum.
The part of the total Hamiltonian, that couples with the external magnetic field, which is assumed to be perpendicular to the bond, is given as
\begin{equation}
    \hat{H}_B= \vec{B}\cdot \hat{L}_\alpha - \chi_{\alpha\alpha} \vec{B}^2= \sum_i \vec{B}\cdot \hat{l}_i - \chi_{\alpha\alpha}\vec{B}^2
\end{equation}
where $\hat{L_\alpha}$ is the total angular momentum operator and $\hat{l_i}$ is the single particle angular momentum operator. To demonstrate this model, the authors consider a closed shell system of 6 electrons such as BH or the CH$^+$ ion. The symmetric ground state is therefore written as 
\begin{equation}
    \Psi_0= |\phi_1\bar{\phi_1}\phi_2\bar{\phi}_2\phi_H\bar{\phi}_H|
\end{equation}
where the notation is such that wavefunctions indexed with 'H' denote the HOMO and those with 'L' denote the LUMO. The doubly degenerate excited states are defined as
\begin{equation}
    \Psi_{1x/y} = \frac{1}{\sqrt{2}} |\phi_1\bar{\phi}_1\phi_2\bar{\phi}_2 (\phi_H\bar{\phi}_{Lx/y}+\phi_{Lx/y}\bar{\phi}_H|.
\end{equation}
Here, let us suppose the $2p_z$ orbital is occupied and hence the LUMO is comprised of two $2p_x$ and $2p_y$, centered on the carbon or boron atom. The Hamiltonian matrix elements between ground and first excited states can then be formulated as
\begin{equation}
    \langle\Psi_0|\hat{H}|\Psi_{1x/y}\rangle=  \langle\phi_H|\hat{l}_{x/y}\cdot\vec{B}|\phi_{Lx/y}\rangle
\end{equation}
We write $\phi_H$ as a linear combination of atomic orbitals, which contribute to hybridization (1s, 2s and $2p_z$), and make the substitutions $2p_y=p_0, 2p_x= \frac{-i}{\sqrt2}(p_1-p_{-1})$ and $2p_z= \frac{1}{\sqrt{2}}(p_1+p_{-1})$, where $p_n$ are the eigenstates of the angular momentum operator $\hat{l}_y$. These substitutions simplify the above equation to yield the matrix elements,
\begin{equation}
     \langle\Psi_0|\hat{H}|\Psi_{1x}\rangle= \iota\mu B ,\hspace{0.4cm}\langle\Psi_0|\hat{H}|\Psi_{1y}\rangle= 0.
\end{equation}
where $\mu$ is the magnetic dipole moment.
A similar approach is followed for an N-membered carbon ring with one $p_z$ orbital at each site. The candidates considered here by the authors are $C_4H_4, C_6H_6$ and $C_{12}H_{12}$. These conjugated systems possess non-degenerate HOMO and LUMO, which correspond to  rotationally allowed eigenstates (eigenstates of $\hat{l}_z$). The HOMO and LUMO are written as a linear superposition of the eigenstates of the single particle angular momentum operator. 
\begin{equation}
    \Psi_{HOMO}= \frac{1}{\sqrt{2}} (|+n\rangle + |-n\rangle), \hspace{0.4cm}   \Psi_{LUMO}= \frac{-\iota}{\sqrt{2}} (|+n\rangle - |-n\rangle)
\end{equation}
Upon calculating the Hamiltonian Matrix elements, a two level Hamiltonian can be constructed as follows,
\begin{equation}
    \hat{H} = \left(\begin {matrix} -\Delta-\chi_0B^2 && \iota\mu B \\ -\iota\mu B && \Delta-\chi_1B^2
 \end {matrix}\right)
\end{equation}
and the eigenvalues obtained upon diagonalizing the Hamiltonian yield energies of the ground and excited state as described in Eq.\ref{2level_energy}.
\begin{equation}
    E_{0/1}= \frac{-1}{2} (\chi_0+\chi_1)B^2 \mp \frac{1}{2}\sqrt{(2\delta+(\chi_0-\chi_1)B^2)^2+4\mu^2B^2}
    \label{2level_energy}
\end{equation}
where $\chi_{0/1}$ are expectation values of the ground and excited state magnetizability tensors, $\Delta$ is the energy difference between the two states in the field-free case. 
It is then possible to write conditions for the magnetic phase transitions. From the definition of magnetizibility, the magnetic phase transition occurs when $\frac{\partial^2E_{0/1}}{\partial B^2}= 2\chi_{0/1}\pm\frac{\mu^2}{4}$ changes its sign. Similarly, the critical magnetic field for magnetic phase transition could be obtained from the condition $\frac{\partial E_{0/1}}{\partial B}=0$. The critical magnetic field for such a phase transition (magnetic field at which paramagnetic to diamagnetic transition occurs) is given as
\begin{equation}
    B_c= \pm \frac{\sqrt{-2\chi_0\chi_1 [\mu^2 + \Delta t]+|\mu(\chi_0+\chi_1)|\sqrt{\chi_0\chi_1(\mu^2+2\Delta t}}}{\chi_0\chi_1t^2}
\end{equation}
where $t=\chi_0-\chi_1$. Further, imposing the condition for the existence of a critical magnetic field, we get a simple condition
\begin{equation}
    \frac{\mu^2}{2\Delta} > |\chi_0|.
\end{equation}
The authors also conclude from this model that the paramagnetic ground state and diamagnetic excited stated only cross when B$\rightarrow$ $\infty$, that is to say that such a crossing is only possible at ultrahigh magnetic fields, not the ones realizable on Earth. In addition, a limitation of this approach is that it does not take into account the energy changes due change in geometry of molecule at high fields.\\
To demonstrate the accuracy of the two level model, the authors compare exact magnetizability (perpendicular field), obtained from linear response function, with that obtained from a eighth order polynomial fit and the two level model for the four molecules, as tabulated below (Table \ref{chi_comparison}). We can clearly see that the two level model is more accurate fit than the eighth order polynomial fit for all the four molecules.

\begin{table}[ht]
\caption{Quantitative comparison of magnetizability($\chi_{\perp}^*$) fitted with the two-level model and the eighth order polynomial. The value of magnetizability obtained from linear response theory is the reference.}
\centering
\begin{tabular}{cccc}
\hline \hline
     \textbf{Species} &\hspace{0.2cm}    \textbf{Linear Response}         &\hspace{0.2cm}  \textbf{Two-Level Model}   &\hspace{0.2cm}                \textbf{8$^{th}$ order polynomial} \\
             BH     &          7.154 &                        7.151 &                                    7.100 \\
       
           $CH^{+}$    &        10.330&                        10.315&                                    10.218\\
     
            $C_{16}H_{10}^{2-}$ & 38.398&                        38.394&                                    38.342\\
            
            $C_{20}H_{10}^{2-}$ & 70.419&                        70.419&                                    70.253\\
    \hline        
    \end{tabular}
    
    \vspace{0.05cm}
     $^*$Note: The subscript ``$\perp$" describes the orientation of the molecule with respect to the magnetic field
    \label{chi_comparison}
\end{table}

\subsection{Coupled Cluster Theory and Density Functional Theory}
In most cases discussed above, the electron-electron correlation was ignored. In this section, methods like the coupled cluster theory and density functional theory are discussed in the context of molecules in high magnetic field. \\
Stopkowicz, et al\cite{Stopkowicz2015} proposed the Coupled Cluster theory for calculating correlation energy in molecules subjected to intense magnetic fields. The authors write the ``coupled cluster" wavefunction as
\begin{equation}
    |\Psi_{cc}\rangle= e^{\hat{T}}|\Phi_0\rangle 
    \label{coupled_cluster_wfc}
\end{equation}
where, $\hat{T}$ is the cluster operator defined as 
\begin{equation}
    \hat{T}= \hat{T}_1 + \hat{T}_2 +... = \sum_{n=1}^N (\frac{1}{n!})^2 \sum_{i,j,k... , a,b,c...}  t_{ij}^{ab}\hat{a}_a^\dagger\hat{a}_b\hat{a}_i^\dagger\hat{a}_j
    \label{cluster_op}
 \end{equation}
where, \textit{a,b} and \textit{i,j} are indices for the virtual and occupied orbitals respectively. $|\Phi_0\rangle$ is chosen as Hartree-Fock wavefunction. Therefore, the correlation energy is computed as the Hartree-Fock expectation value.
\begin{equation}
    E_{corr}^{cc}= \langle\Psi_{HF}|e^{-\hat{T}}\hat{H}_Ne^{\hat{T}}|\Psi_{HF}\rangle
\end{equation}
That is to say that the total Hamiltonian is the sum of Hartree-Fock energy ($E_{HF}$) and the coupled cluster Hamiltonian ($H_N$), which describes correlation effects.

\begin{figure}[ht]
    \centering
    \includegraphics[width=0.5\textwidth]{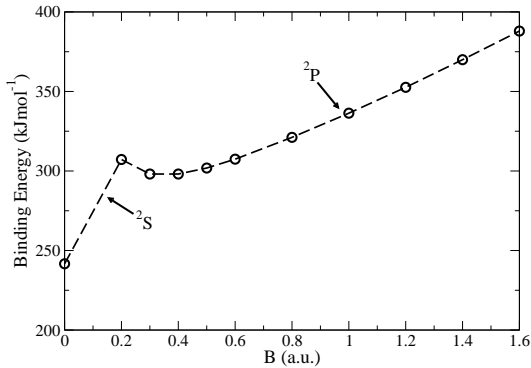}
    \caption{Binding energy of LiH in the perpendicular orientation. The $^2S$ state is stable till B=0.2 a.u. but the $^2P$ state is lower in energy for $B>0.2 a.u.$ \textit{Data adapted from \cite{Stopkowicz2015}.}}
    \label{LiH_paramagnetic}
\end{figure}

This model is then applied to the simplest case where correlation effects could be studied; the Helium atom. In this case, the diamagnetic to paramagnetic crossover is observed at approximately $0.8 a.u.$ of magnetic field. The authors show that the energy of the diamagnetic singlet state increases continuously with magnetic field, while that of the triplet paramagnetic state decreases. This is linked by the authors to the spatial extension of orbitals , which destablilize the singlet. They present the argument by comparing the expectation value of the diamagnetic contribution (Eq.\ref{diamagnetic_contri}), in the correlated regime, to that of the Hartree-Fock result. 
\begin{equation}
    E_{dia}= \frac{1}{8}B_z^2\langle\Psi|\hat{r}^2|\Psi\rangle 
    \label{diamagnetic_contri}
\end{equation}
This quantity, for HF level of theory, is lower as compared to the CCSD (Coupled Cluster Singles and Doubles) theory. Qualitatively, it means that there is more correlation in the singlet state, where we have 2 electrons in the 1s orbital, whereas the correlation is lowered upon promoting one electron to the 2p orbital, in case of the triplet. In addition, the correlation in the 1s would cause expansion of the orbital, reducing the screening for the 2p orbital. This causes spatial contraction in the 2p orbital. However, a general trend is not observed; for instance the singlet and triplet states show greater spatial extension in case of Hartree-Fock while the singlet of Ne shows greater spatial extension in case of coupled cluster theory calculations.  
The authors also report the case of LiH molecule. The paramagnetic bonding in LiH is interpreted in two ways; by noting the destabilization of the $^1\Sigma_\|$ and stabilization of $^1\Sigma_\bot$ with increasing magnetic field. Secondly, the binding energy profile for LiH in Fig.\ref{LiH_paramagnetic} shows continuous increase in the binding energy for magntic field strength greater than 0.2 a.u. Reconciliation of these two observations confirm paramgnetic bonding in LiH.

Around the same time, in 2015, Furness et al\cite{Furness2015} devised a way to determnine magnetic response of molecules using density functional theory, known as the Current Density Functional Theory (CDFT). The authors introduce a formalism that uses the meta-GGA (Meta-Generalized Gradient Approximation) exchange-correlation. Meta-GGA (MGGA) functionals have proven to be very successful in predicting effects of correlation, and therefore could potentially be exploited to address the problems posed by HF based methods. However, traditional MGGA functionals depend on canonical kinetic energy operator, in which case, gauge invariance does not hold. In order to circumvent this issue, the authors replace the canonical kinetic energy by its generalized form including the paramagnetic current density.\\
The mathematical formulation of CDFT is compactly written as in Eq.\ref{cdft}
\begin{equation}
    [\frac{\hat{p}^2}{2m}+\frac{1}{2m} {\hat{p},\hat{A_s}}+V_{ext}+ \frac{1}{2}\hat{A}_{ext}^2+ V_C+ V_xc+\hat{s}\cdot(\hat{\nabla}\times \hat{A}_s)] \psi_i = \varepsilon_i \psi_i
    \label{cdft}
\end{equation}
Here, $A_s$ is the sum of external vector potential ($A_{ext}$ and the one due to exchange-correlation interaction ($A_{xc}$. It is important to distinguish between the latter and $V_{xc}$. They are defined, as a function of electron density ($\rho$) and current density ($j$) in the following way:
\begin{equation}
    V_{xc}= \frac{\delta E_{xc}(\rho,j)}{\delta\rho(\vec{r})}, \hspace{0.5cm} A_{xc}= \frac{\delta E_{xc}(\rho,j)}{\delta j(\vec{r})}.
\end{equation}
The aforementioned Kohn-Sham system is solved iteratively using the uncontracted aug-cc-p-CVTZ basis set(citation needed). The authors use the cTPSS, cTPSS(h), B98 and the KT3(Keal-Tozer-3) functionals, available with the software package LONDON. Success of these MGGA functionals is decided by comparison with the methods established before (see Table\ref{dft_comparison}).

\begin{table}[ht]
\caption{Comparison of different levels of theory used to predict paramagnetic bonding in $H_2$ and $He_2$. The quality of results for H$_2$ and $He_2$ molecules is examined by comparison with the FCI calculations, and the CCSD(T) calculations respectively. }

\centering
\begin{tabular}{ccccc}
\hline \hline
     \textbf{Species}& \textbf{Hartree-Fock}         &  \textbf{LDA}                  &\textbf{ PBE} &    \textbf{          cTPSS/cTPSS(h)}  \\
             $H_2$   & strongly underbinds  &  strongly overbinds   & better than LDA   & closest to the FCI description\\
    \hline
             $He_2$  & strongly underbinds   & strongly overbinds and      & Overbinds, but   & good description of potential energy\\
                     & and overestimates     & gives too short             & better than LDA. & but bond lengths and diss. energy\\
                     & bond length.          & bond length.                & (also for KT3)   & suggest strong tendency to overbind.\\
    \hline  
    \end{tabular}
    \label{dft_comparison}
\end{table}
In case of the H$_2$ molecule, the KT3 functional is found to yield reasonably accurate number for the dissociation energy and bond length, but an absurd barrier in potential energy is observed. The barrier, in case of B98, is even bigger. Hence, these functionals, although, give reasonable magnitudes cannot be considered reliable to extract the physics of $H_2$. In the case of He$_2$, the B98 functional underbinds,  but since it is empirically parametrized, the authors claim that its optimization could be enhanced. 

\begin{figure*} [ht]
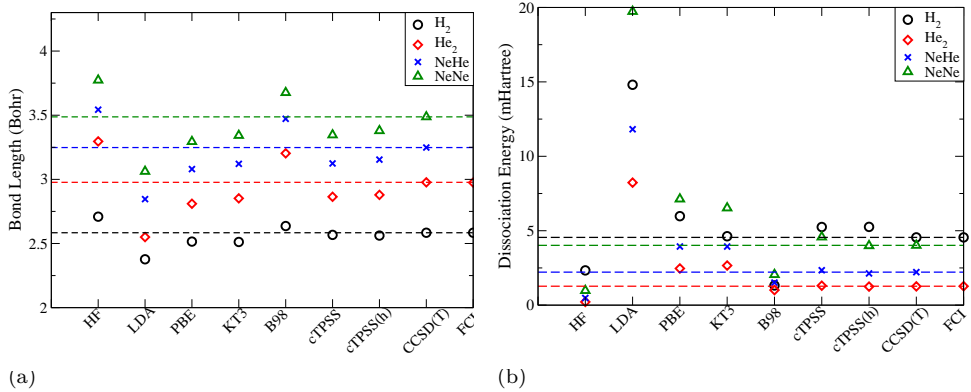

 \centering
    \begin{subfigure}[]
    \centering
    \includegraphics[width=0.45\textwidth]{cdft_comp.eps}
    \label{fig:fig2a}
    \end{subfigure}
    \begin{subfigure}[]
    \centering
    \includegraphics[width=0.45\textwidth]{diss_energy_cdft.eps}
    \end{subfigure}
\caption{Performance of different functionals/levels of theory to compute\textbf{(a)} bond length and \textbf{(b)}dissociation energies of $H_2, He_2$, NeHe and NeNe in the presence of 1 a.u. magnetic field. The horizontal dashed lines are drawn as the reference (FCI or CCSD(T)) for other functionals (HF, LDA, PBE, KT3, B98, cTPSS and cTPSS(h)}
\label{cdft_comparison}
\end{figure*}
Similar trends are observed in the case of NeHe and NeNe, and are summarised in Fig\ref{cdft_comparison}.
The authors report on the paramagnetic bonding using different functionals and levels of theory, as a measure to benchmark the performance of meta-GGA functionals. This is carried out for $H_2$, $He_2$, HeNe cluster and NeNe dimer. The paramagnetic bonding is explored by computing the variation in bond lengths. The authors also correctly predict perpendicular paramagnetic bonding (by examination of bond lengths and electron density plots). 

Among the more recent calculation methods, Equation of Motion CCSD (EOM-CCSD) was introduced by Hampe et al\cite{Hampe2017}, in 2017. In this method, the authors assume a general excited state wavefunction generated by the action of the action of a general excitation operator $\hat{R}$, which assumes the same form as that of the cluster operator, defined in Eq.\ref{cluster_op}.
\begin{equation}
    \hat{R}= \hat{R}_1 + \hat{R}_2 +... = \sum_{n=1}^N (\frac{1}{n!})^2 \sum_{i,j,k... , a,b,c...}  r_{ij}^{ab}\hat{a}_a^\dagger\hat{a}_b\hat{a}_i^\dagger\hat{a}_j
    \label{exc_op}
 \end{equation}
 The energy is also computed in the same way,
 \begin{equation}
     e^{-\hat{T}}(\hat{H}-E_{cc})e^{\hat{T}} \hat{R}|\phi_0\rangle = \Delta E_{exc}\hat{R}|\phi_0\rangle
 \end{equation}
 Here, $\Delta E_{exc}$ is the difference between energy of the excited and reference coupled cluster state.\\
 The authors discuss two kinds of excitations in their article; electronic and spin-flip. In case of the former, the excitation occurs in such a way that the total spin of the system is conserved. However, in case of spin-flip excitation, the total spin of the system may not be conserved. Hence, for an overall spin-less system, a spin-flip excitation can give rise to an overall spin 1. 
 
 The authors consider C, $H_2$ and $CH^+$ ion in magnetic fields upto 1a.u. For carbon, they consider $1s^22s2p_{-1}2p_02p_{+1}$ as the reference state, while it is found that the excited states have the configuration $1s^22s^22p_{-1}2p_0 (^3\Pi_g)$, $1s^22s2p_{-1}^22p_02 (^3\Delta_u)$, $1s^22s2p_{-1}2p_03d_{-2} (^5\Phi_g)$. The authors compute the magnetic field at which the state crossovers happen and compare with Ivanov et al\cite{Ivanov1999}. Their results are tabulated in Table\ref{carbon_eom-ccsd}.
 \begin{table}[ht]
   \caption{Magnetic Field strengths for configuration crossover computed using EOM-CCSD are compared with corresponding Hartree-Fock values reported in \cite{Ivanov1999}. All the magnetic fields reported are in atomic units.}
     \centering

     \begin{tabular}{ccc}
     \hline \hline
    \textbf{Configuration}  & \textbf{B}\cite{Ivanov1999} & \textbf{B}(EOM-CCSD)  \\
          $^3\Pi_g$             & $\approx$ 0.1862              & $\approx$0.313      \\
          $^5\Sigma_u$          & 0.1862-0.4903               & 0.313-0.513       \\
          $^5\Phi_g$            & B $>$ 0.4903                & B $>$ 0.513     \\
           \hline
     \end{tabular}
     \label{carbon_eom-ccsd}
 \end{table}
 From the above observations, it is clear that electron-electron correlation also influences crossover magnetic field strengths. However, in the case of $H_2$ molecule, the authors find no significant difference between the FCI and EOM-CCSD calculations. For the $CH^+$ ion, a similar result is obtained where $^1\Delta$ is the lowest lying state in high magnetic field, parallel to the molecule. \\
 Therefore, not only does this method of calculation consider correlation effects to a reasonable accuracy, but it is also ovrecomes the limitations of a FCI calculations, of applicability to large systems. 
 
 \section{Summary }
In this review we began by presenting an overview of atoms and molecules in weak magnetic field. We discussed the problem of gauge invariance, which is also shared by the strong field case. We briefly discussed the GIPAW and CTOCD-DZ methods used by most softwares to overcome this problem.\\
To the best of our knowledge, a systematic computational approach towards calculating energy spectra of atoms in high magnetic field was presented by Kappes in 1994. Although, the Hartree-Fock theory for the same existed even before. The primary interest of all theoretical calculations was to study the bonding behavior of molecules (of astrophysical interest) in ultrahigh magnetic fields. Hartree-Fock calculations were successful in predicting most of the ground state configurations in high magnetic field regimes and also predicted paramgnetic bonding in $H_2$, qualitatively. \\
We also discussed a two-level model for rationalizing Hartree-Fock results for paramagnetic closed shell molecules, which is fits more accurately, that the $8^{th}$ order polynomial fit (derived out of Taylor expansion), to the HF data. 
The shortcomings of Hartree-Fock were overcome by FCI, CCSD (/EOM-CCSD) and DFT methods, where electron-electron correlation is not ignored. However, FCI cannot be considered a suitable method for most practical cases, where the system size is very large. Nonetheless, works of Stopkowicz et al, Furness et al and Hampe et al show that CCSD and DFT (with meta-GGA functionals) are extremely successful for systems larger than the He atom. However, recent observations indicate significant abundance of transition metals/minerals\cite{Jura2014} and therefore it needs to be tested how the aforementioned methods perform for larger systems. Lastly, to the best of our knowledge, no models have been formulated for systems having a diamagnetic ground state. Therefore, we propose it to be a worthwhile attempt to rationalize computational data for these systems based on simple theoretical models like the one discussed in Section \ref{strong_field_sec}C.

\bibliographystyle{plain}
\bibliography{full_bib6}

\begin{thebibliography}{10}

\bibitem{Aidas2014}
Kestutis Aidas, Celestino Angeli, Keld~L. Bak, Vebj{\o}rn Bakken, Radovan Bast,
  Linus Boman, Ove Christiansen, Renzo Cimiraglia, Sonia Coriani, P{\aa}l
  Dahle, Erik~K. Dalskov, Ulf Ekstr{\"{o}}m, Thomas Enevoldsen, Janus~J.
  Eriksen, Patrick Ettenhuber, Berta Fern{\'{a}}ndez, Lara Ferrighi, Heike
  Fliegl, Luca Frediani, Kasper Hald, Asger Halkier, Christof H{\"{a}}ttig,
  Hanne Heiberg, Trygve Helgaker, Alf~Christian Hennum, Hinne Hettema, Eirik
  Hjerten{\ae}s, Stinne H{\o}st, Ida~Marie H{\o}yvik, Maria~Francesca Iozzi,
  Branislav Jans{\'{i}}k, Hans J{\o}rgen~Aa Jensen, Dan Jonsson, Poul
  J{\o}rgensen, Joanna Kauczor, Sheela Kirpekar, Thomas Kj{\ae}rgaard, Wim
  Klopper, Stefan Knecht, Rika Kobayashi, Henrik Koch, Jacob Kongsted, Andreas
  Krapp, Kasper Kristensen, Andrea Ligabue, Ola~B. Lutn{\ae}s, Juan~I. Melo,
  Kurt~V. Mikkelsen, Rolf~H. Myhre, Christian Neiss, Christian~B. Nielsen,
  Patrick Norman, Jeppe Olsen, J{\'{o}}gvan Magnus~H. Olsen, Anders Osted,
  Martin~J. Packer, Filip Pawlowski, Thomas~B. Pedersen, Patricio~F. Provasi,
  Simen Reine, Zilvinas Rinkevicius, Torgeir~A. Ruden, Kenneth Ruud,
  Vladimir~V. Rybkin, Pawel Sa{\l}ek, Claire~C.M. Samson, Alfredo~S{\'{a}}nchez
  de~Mer{\'{a}}s, Trond Saue, Stephan~P.A. Sauer, Bernd Schimmelpfennig,
  Kristian Sneskov, Arnfinn~H. Steindal, Kristian~O. Sylvester-Hvid, Peter~R.
  Taylor, Andrew~M. Teale, Erik~I. Tellgren, David~P. Tew, Andreas~J.
  Thorvaldsen, Lea Th{\o}gersen, Olav Vahtras, Mark~A. Watson, David~J.D.
  Wilson, Marcin Ziolkowski, and Hans {\AA}gren.
\newblock {The Dalton quantum chemistry program system}.
\newblock {\em Wiley Interdisciplinary Reviews: Computational Molecular
  Science}, 4(3):269--284, 2014.

\bibitem{Boblest2014}
Sebastian Boblest, Christoph Schimeczek, and G{\"{u}}nter Wunner.
\newblock {Ground states of helium to neon and their ions in strong magnetic
  fields}.
\newblock {\em Physical Review A - Atomic, Molecular, and Optical Physics},
  89(1):1--8, 2014.

\bibitem{Clark2005}
Stewart~J. Clark, Matthew~D. Segall, Chris~J. Pickard, Phil~J. Hasnip,
  Matt~I.J. Probert, Keith Refson, and Mike~C. Payne.
\newblock {First principles methods using CASTEP}.
\newblock {\em Zeitschrift fur Kristallographie}, 220(5-6):567--570, 2005.

\bibitem{Demeur1994}
M.~Demeur, P.~H. Heenen, and M.~Godefroid.
\newblock {Hartree-Fock study of molecules in very intense magnetic fields}.
\newblock {\em Physical Review A}, 49(1):176--183, 1994.

\bibitem{Fowler1991}
P~W Fowler and E~Steiner.
\newblock {Paramagnetic closed-shell molecules: the isoelectronic series CH+,
  BH and BeH-}.
\newblock {\em Molecular Physics}, 74(6):1147--1158, 1991.

\bibitem{Fowler1996}
P.~W. Fowler, R.~Zanasi, B.~Cadioli, and E.~Steiner.
\newblock {Ring currents and magnetic properties of pyracylene}.
\newblock {\em Chemical Physics Letters}, 251(3-4):132--140, 1996.

\bibitem{Furness2015}
James~W. Furness, Joachim Verbeke, Erik~I. Tellgren, Stella Stopkowicz, Ulf
  Ekstr{\"{o}}m, Trygve Helgaker, and Andrew~M. Teale.
\newblock {Current Density Functional Theory Using Meta-Generalized Gradient
  Exchange-Correlation Functionals}.
\newblock {\em Journal of Chemical Theory and Computation}, 11(9):4169--4181,
  2015.

\bibitem{Hammond1994}
B.~L. Hammond, W.~A.~Jr Lester, and P.~J. Reynolds.
\newblock {\em {Monk Carb Methods in Ab Iniio Quantum Chemistry}}, volume~1.
\newblock World Scientific, 1994.

\bibitem{Hampe2017}
Florian Hampe and Stella Stopkowicz.
\newblock {Equation-of-motion coupled-cluster methods for atoms and molecules
  in strong magnetic fields}.
\newblock {\em Journal of Chemical Physics}, 146(15), 2017.

\bibitem{Hegstrom1968a}
R.~A. Hegstrom and W.~N. Lipscomb.
\newblock {Magnetic properties of the BF and BH molecules}.
\newblock {\em The Journal of Chemical Physics}, 48(2):809, 1968.

\bibitem{Hegstrom1966}
Roger~A. Hegstrom and William~N. Lipscomb.
\newblock {Magnetic properties of the BH molecule}.
\newblock {\em The Journal of Chemical Physics}, 45(7):2378--2383, 1966.

\bibitem{Hegstrom1968}
Roger~A. Hegstrom and William~N. Lipscomb.
\newblock {Paramagnetism in closed-shell molecules}.
\newblock {\em Reviews of Modern Physics}, 40(2):354--358, 1968.

\bibitem{Helgaker1999}
Trygve Helgaker, Michal Jaszu{\'{n}}ski, and Kenneth Ruud.
\newblock {Ab initio methods for the calculation of NMR shielding and indirect
  spin-spin coupling constants}.
\newblock {\em Chemical Reviews}, 99(1):293--352, 1999.

\bibitem{Ho2009}
Wynn~C.G. Ho and Craig~O. Heinke.
\newblock {A neutron star with a carbon atmosphere in the Cassiopeia A
  supernova remnant}.
\newblock {\em Nature}, 462(7269):71--73, 2009.

\bibitem{Ivanov1999}
M.~V. Ivanov and P.~Schmelcher.
\newblock {Ground state of the carbon atom in strong magnetic fields}.
\newblock {\em Physical Review A - Atomic, Molecular, and Optical Physics},
  60(5):3558--3568, 1999.

\bibitem{Ivanov2000}
M.~V. Ivanov and P.~Schmelcher.
\newblock {Ground states of H, He,{\ldots}, Ne, and their singly positive ions
  in strong magnetic fields: The high-field regime}.
\newblock {\em Physical Review A - Atomic, Molecular, and Optical Physics},
  61(2):13, 2000.

\bibitem{Ostriker1968}
F.~D. A.~Ostriker {Jeremiah P.} and Hartwick.
\newblock {RAPIDLY ROTATING STARS. IV. MAGNETIC WHITE DWARFS}.
\newblock {\em The Astrophysical Journal}, 153:797--806, 1968.

\bibitem{Joyce2007}
Si{\^{a}}n~A. Joyce, Jonathan~R. Yates, Chris~J. Pickard, and Francesco Mauri.
\newblock {A first principles theory of nuclear magnetic resonance J -coupling
  in solid-state systems}.
\newblock {\em Journal of Chemical Physics}, 127(20), 2007.

\bibitem{Jura2014}
M.~Jura and E.D. Young.
\newblock {Extrasolar Cosmochemistry}.
\newblock {\em Annual Review of Earth and Planetary Sciences}, 42(1):45--67,
  2014.

\bibitem{Kappes1994}
U.~Kappes and P.~Schmelcher.
\newblock {Atomic orbital basis set optimization for ab initio calculations of
  molecules with hydrogen atoms in strong magnetic fields}.
\newblock {\em The Journal of Chemical Physics}, 100(4):2878--2887, 1994.

\bibitem{Keith1993}
Todd~A. Keith and Richard~F.W. Bader.
\newblock {Calculation of magnetic response properties using a continuous set
  of gauge transformations}.
\newblock {\em Chemical Physics Letters}, 210(1-3):223--231, 1993.

\bibitem{Lange2012}
Kai~K. Lange, E.~I. Tellgren, M.~R. Hoffmann, and T.~Helgaker.
\newblock {A paramagnetic bonding mechanism for diatomics in strong magnetic
  fields}.
\newblock {\em Science}, 337(6092):327--331, 2012.

\bibitem{Laskowski2012}
Robert Laskowski and Peter Blaha.
\newblock {Calculations of NMR chemical shifts with APW-based methods}.
\newblock {\em Physical Review B - Condensed Matter and Materials Physics},
  85(3), 2012.

\bibitem{Laskowski2014}
Robert Laskowski and Peter Blaha.
\newblock {Calculating NMR chemical shifts using the augmented plane-wave
  method}.
\newblock {\em Physical Review B - Condensed Matter and Materials Physics},
  89(1):1--7, 2014.

\bibitem{Lazzeretti1994}
Paolo Lazzeretti, Massimo Malagoli, and Riccardo Zanasi.
\newblock {Computational approach to molecular magnetic properties by
  continuous transformation of the origin of the current density}.
\newblock {\em Chemical Physics Letters}, 220(3-5):299--304, 1994.

\bibitem{Lazzeretti1995}
Paolo Lazzeretti, Massimo Malagoli, Riccardo Zanasi, Ernest~W. Delia, Ian~J.
  Lochert, Claudia~G. Giribet, Mart{\'{i}}n~C.Ruiz {De Az{\'{u}}a}, and
  Rub{\'{e}}n~H. Contreras.
\newblock {Ab initio and experimental study of NMR coupling constants in
  bicyclo[1.1.1]pentane}.
\newblock {\em Journal of the Chemical Society, Faraday Transactions},
  91(22):4031--4035, 1995.

\bibitem{Lazzeretti1976}
Paolo Lazzeretti, Ferdinando Taddei, and Riccardo Zanasi.
\newblock {Coupled Hartree-Fock Calculations of Nuclear Magnetic Resonance
  Carbon-Carbon Coupling Constants in Substituted Benzenes}.
\newblock {\em Journal of the American Chemical Society}, 98(25):7989--7993,
  1976.

\bibitem{Lazzeretti1980}
Paolo Lazzeretti and Riccardo Zanasi.
\newblock { Theoretical determination of the magnetic properties of HCl, H 2 S,
  PH 3 , and SiH 4 molecules }.
\newblock {\em The Journal of Chemical Physics}, 72(12):6768--6776, 1980.

\bibitem{Ligabue2003}
Andrea Ligabue, Stephan~P.A. Sauer, and Paolo Lazzeretti.
\newblock {Correlated and gauge invariant calculations of nuclear magnetic
  shielding constants using the continuous transformation of the origin of the
  current density approach}.
\newblock {\em Journal of Chemical Physics}, 118(15):6830--6845, 2003.

\bibitem{McMurchie1978}
Larry~E. McMurchie and Ernest~R. Davidson.
\newblock {One- and two-electron integrals over cartesian gaussian functions}.
\newblock {\em Journal of Computational Physics}, 26(2):218--231, 1978.

\bibitem{Neuhauser1986}
D.~Neuhauser, K.~Langanke, and S.~E. Koonin.
\newblock {Hartree-Fock calculations of atoms and molecular chains in strong
  magnetic fields}.
\newblock {\em Physical Review A}, 33(3):2084--2086, 1986.

\bibitem{Pickard2001}
Chris~J. Pickard and Francesco Mauri.
\newblock {All-electron magnetic response with pseudopotentials: NMR chemical
  shifts}.
\newblock {\em Physical Review B - Condensed Matter and Materials Physics},
  63(24):2451011--2451013, 2001.

\bibitem{Ramsey1951}
Norman~F. Ramsey.
\newblock {Magnetic shielding of nuclei in molecules}.
\newblock {\em Physica}, 17(3-4):303--307, 1951.

\bibitem{Ruder1984}
H.~Ruder, H.~Herold, W.~R{\"{o}}sner, and G.~Wunner.
\newblock {Pulsars: High magnetic field laboratories with 10$^0$ T}.
\newblock {\em Physica B+C}, 127(1-3):11--25, 1984.

\bibitem{Ruderman1971}
M.~Ruderman.
\newblock {Matter in superstrong magnetic fields: The surface of a neutron
  star}.
\newblock {\em Physical Review Letters}, 27(19):1306--1308, 1971.

\bibitem{Stopkowicz2015}
Stella Stopkowicz, J{\"{u}}rgen Gauss, Kai~K. Lange, Erik~I. Tellgren, and
  Trygve Helgaker.
\newblock {Coupled-cluster theory for atoms and molecules in strong magnetic
  fields}.
\newblock {\em Journal of Chemical Physics}, 143(7), 2015.

\bibitem{Suleimanov2014}
V.~F. Suleimanov, D.~Klochkov, G.~G. Pavlov, and K.~Werner.
\newblock {Carbon neutron star atmospheres}.
\newblock {\em Astrophysical Journal, Supplement Series}, 210(1), 2014.

\bibitem{Tellgren2009}
Erik~I. Tellgren, Trygve Helgaker, and Alessandro Soncini.
\newblock {Non-perturbative magnetic phenomena in closed-shell paramagnetic
  molecules}.
\newblock {\em Physical Chemistry Chemical Physics}, 11(26):5489--5498, 2009.

\bibitem{Tellgren2012}
Erik~I. Tellgren, Simen~S. Reine, and Trygve Helgaker.
\newblock {Analytical GIAO and hybrid-basis integral derivatives: Application
  to geometry optimization of molecules in strong magnetic fields}.
\newblock {\em Physical Chemistry Chemical Physics}, 14(26):9492--9499, 2012.

\bibitem{Tellgren2008}
Erik~I. Tellgren, Alessandro Soncini, and Trygve Helgaker.
\newblock {Nonperturbative ab initio calculations in strong magnetic fields
  using London orbitals}.
\newblock {\em Journal of Chemical Physics}, 129(15):1--11, 2008.

\bibitem{Thirumalai2014}
Anand Thirumalai, Steven~J. Desch, and Patrick Young.
\newblock {Carbon atom in intense magnetic fields}.
\newblock {\em Physical Review A - Atomic, Molecular, and Optical Physics},
  90(5):1--8, 2014.

\bibitem{Vignale1987}
G.~Vignale and Mark Rasolt.
\newblock {Density-functional theory in strong magnetic fields}.
\newblock {\em Physical Review Letters}, 59(20):2360--2363, 1987.

\end{thebibliography}
\end{document}